\documentclass[aip,jap,reprint]{revtex4-1}
\usepackage{bm}
\usepackage{amsmath}
\usepackage[dvipdfmx]{graphicx,color}

\def\Vec#1{\mbox{\boldmath $#1$}}

\def\r2A20{$\langle r^{2} \rangle A_{2}^{0}$}

\def\NdFeCoN{NdFe$_{11}$CoN}
\def\NdFeTiN{NdFe$_{11}$TiN}
\def\NdFeMN{NdFe$_{11}M$N}
\def\NdFeN{NdFe$_{12}$N}
\def\NdFeCo{NdFe$_{11}$Co}
\def\NdFeTi{NdFe$_{11}$Ti}
\def\NdFeM{NdFe$_{11}M$}
\def\NdFe{NdFe$_{12}$}

\begin{document}

\title{First-principles study on stability and magnetism of
  NdFe$_{11}M$ and NdFe$_{11}M$N \\
  for $M$=Ti, V, Cr, Mn, Fe, Co, Ni, Cu, Zn}

%%% affiliation in desired order %%%
\affiliation{Research Center for Computational Design of Advanced Functional Materials, 
  National Institute of Advanced Industrial Science and Technology, 
  Tsukuba, Ibaraki 305-8568, Japan}
\affiliation{Center for Materials Research by Information Integration, 
  National Institute for Materials Science, Tsukuba, Ibaraki 305-0047, Japan}
\affiliation{Elements Strategy Initiative Center for Magnetic Materials, 
  National Institute for Materials Science, Tsukuba, Ibaraki 305-0047, Japan}

%%% authors with affiliation %%%
\author{Yosuke Harashima}
\affiliation{Research Center for Computational Design of Advanced Functional Materials, 
  National Institute of Advanced Industrial Science and Technology, 
  Tsukuba, Ibaraki 305-8568, Japan}
\affiliation{Elements Strategy Initiative Center for Magnetic Materials, 
  National Institute for Materials Science, Tsukuba, Ibaraki 305-0047, Japan}

\author{Kiyoyuki Terakura}
\affiliation{Center for Materials Research by Information Integration, 
  National Institute for Materials Science, Tsukuba, Ibaraki 305-0047, Japan}

\author{Hiori Kino}
\affiliation{Center for Materials Research by Information Integration, 
  National Institute for Materials Science, Tsukuba, Ibaraki 305-0047, Japan}
\affiliation{Elements Strategy Initiative Center for Magnetic Materials, 
  National Institute for Materials Science, Tsukuba, Ibaraki 305-0047, Japan}

\author{Shoji Ishibashi}
\affiliation{Research Center for Computational Design of Advanced Functional Materials, 
  National Institute of Advanced Industrial Science and Technology, 
  Tsukuba, Ibaraki 305-8568, Japan}

\author{Takashi Miyake}
\affiliation{Research Center for Computational Design of Advanced Functional Materials, 
  National Institute of Advanced Industrial Science and Technology, 
  Tsukuba, Ibaraki 305-8568, Japan}
\affiliation{Center for Materials Research by Information Integration, 
  National Institute for Materials Science, Tsukuba, Ibaraki 305-0047, Japan}
\affiliation{Elements Strategy Initiative Center for Magnetic Materials, 
  National Institute for Materials Science, Tsukuba, Ibaraki 305-0047, Japan}

\date{\today}

\begin{abstract}
  Recently synthesized NdFe$_{12}$N has excellent magnetic properties, 
  while it is thermodynamically unstable.
  Using first-principles method, 
  we study the effect of substitutional 3$d$ transition metal elements 
  to the mother compound NdFe$_{12}$. 
  We find that Co has positive effect on the stability of the ThMn$_{12}$ structure. 
  In contrast with Ti substitution, Co substitution does not 
  reduce the magnetization significantly.
  The crystal field parameter $\langle r^{2} \rangle A_{2}^{0}$ is nearly unchanged by Co substitution, and 
  nitrogenation to NdFe$_{11}$Co greatly enhances $\langle r^{2} \rangle A_{2}^{0}$. 
  This suggests that Co is a good candidate as 
  a substitutional element for NdFe$_{12}$N.
\end{abstract}

\maketitle

%%% Introduction %%%
\section{Introduction} \label{sec:introduction}
Recently, NdFe$_{12}$N has been synthesized \cite{HiTaHiHo2015} following suggestion 
by a theoretical work \cite{MiTeHaKiIs2014}.
It exhibits larger magnetization and
stronger uniaxial magnetocrystalline anisotropy than Nd$_{2}$Fe$_{14}$B
at and above the room temperature.
NdFe$_{12}$N is one of 
ThMn$_{12}$-type rare-earth transition-metal alloys,
where Nd atoms occupy the Th sites and Fe atoms occupy the Mn sites.
The crystal structure contains high ratio of transition-metal sites to rare-earth sites. 
It is favorable for achieving large magnetization, 
hence iron-based ThMn$_{12}$-type compounds have been studied 
as potential candidates for permanent magnet materials.
\cite{SuKuUrKoSaWaKiKaMa2014,HaTeKiIsMi2015a,HiMiHo2015,
  SuKuUrKoSaWaYaKaMa2016,KuSuUrKoSaYaKaMa2016,KoKrEl2016,KeJo2016,HaFuTeKiIsMi2016}
It was reported in the late 80's that 
SmFe$_{11}$Ti has reasonably large magnetization and magnetocrystalline anisotropy. 
Although NdFe$_{11}$Ti does not show uniaxial anisotropy, 
interstitial nitrogenation induces strong uniaxial anisotropy, and also 
increases the magnetization. 
\cite{YaZhKoPaGe1991,YaZhGePaKoLiYaZhDiYe1991}

The iron-based ThMn$_{12}$-type compounds are normally synthesized without nitrogen.
Nitrogen is doped afterwards into the interstitial sites of the mother compound 
to improve magnetic properties, if necessary.
In the step synthesizing the mother compound, 
it has been known that $R$Fe$_{12}$ is unstable thermodynamically.
For applications to permanent magnets, a bulk sample is necessary.
However, the synthesis of $R$Fe$_{12}$ has been succeeded
only as a film sample.\cite{CaHeNaRaCh1991,WaLiHeSeHaZh1993,HiTaHiHo2015}
A third element is necessary to stabilize
the ThMn$_{12}$-type crystal structure as a bulk sample.
Thus far, several elements are known to stabilize the structure. 
In addition to titanium, V, Cr, Mn, Mo, W, Al, Si serve as stabilizing elements 
when they substitute the Fe sites. 
\cite{Fe1980,YaKeJaDeYe1981,MoBu1988,OhTaOsSaKo1988,Mu1988,WaChBeEtCoCa1988}
The concentration range for stability depends on the element
(summarized in Fig.~1 of Ref.~\onlinecite{Co1990}).

The substitution reduces the Fe content, hence 
the magnetization of $R$Fe$_{12-x}M_{x}$ is normally smaller than that of $R$Fe$_{12}$.
For example, Ti substitution does reduce the magnetization. 
In fact, the magnetization is even smaller than 
naive expectation from the iron concentration. \cite{MiTeHaKiIs2014,HaFuTeKiIsMi2016}
This substantial reduction is explained qualitatively 
by Friedel's concept of virtual bound state.
\cite{Fr1958,VeBoZhBu1988,MiTeHaKiIs2014}
Therefore, 
a substitutional element that stabilizes the crystal structure without 
substantial magnetization reduction is required for 
applications to permanent magnets.

In this paper, we give systematic research to find 
better substitutional elements in terms of the stability, magnetization, 
and magnetocrystalline anisotropy in \NdFeMN. 
Our scheme is as follows: 
We first study the stability of \NdFeM\ 
where $M$ is a 3$d$ transition-metal element from Ti to Zn. 
We also estimate overall behavior of magnetic moments. 
Next we will study their nitrogenated crystal to calculate
magnetic moments and crystal field parameter $\langle r^{2} \rangle A_{2}^{0}$ for good candidate(s).
We follow experimental synthesis process, 
i.e., synthesizing the mother compounds \NdFeM\ first, then, 
nitrogenating the mother compounds later.
Calculated data will be also useful to compare the properties between 
our theory and experiments at each step.
All of these works are carried out in the first-principles calculations.

%%% Calculation methods %%%
\section{Calculation methods} \label{sec:methods}
We perform the first-principles calculations by using 
QMAS (Quantum MAterials Simulator)~\cite{Qm2014} 
which is based on the density functional theory~\cite{HoKo1964,KoSh1965} 
and the projector augmented-wave method.\cite{Bl1994,KrJo1999} 
The exchange-correlation energy functional is given by the PBE formula 
in the generalized gradient approximation.~\cite{PeBuEr1996}
We use the open-core approximation for the Nd-4$f$, 
in which Nd-4$f$ is not treated as the valence states 
(see Ref.~\onlinecite{Ri1998} and references therein). 
All the calculations are done in 
the collinear spin alignment and without the spin-orbit coupling. 
After the calculation, we approximate the total magnetic moment by hand 
by adding the total spin magnetic moment of the valence electrons 
with magnetic moment of the Nd-4$f$ open-core contribution 
$g_{J}J$=3.273 $\mu_{B}$, where $g_{J}$ is the Lande g-factor and 
$J$ is the total angular momentum $9/2$ as given by Hund's rules. 
Note that this approximation does not include either the effects of 
(A) the orbital magnetic moments of other elements (e.g, Fe, Co) 
or (B) the hybridization of Nd-4$f$ with the other orbitals. 

In this study, we consider the magnetocrystalline anisotropy 
based on the crystal-field theory. 
Within the lowest order approximation based on this theory,
contribution of the Nd-4$f$ electrons to 
the magnetocrystalline anisotropy constant is given by 
\begin{equation} \label{eq:mae-cef}
  K_{1} = -3J(J-\frac{1}{2}) \alpha_{J} \langle r^{2} \rangle A_{2}^{0} n_{R} \;,
\end{equation}
where $n_{R}$ is a concentration of Nd atoms and
$\alpha_{J}$ is the Stevens factor, $-7/1089$ for Nd$^{3+}$.
$\langle r^{2} \rangle A_{2}^{0}$ is the crystal field parameter calculated as,
\begin{equation} \label{eq:cef_veff}
  \langle r^{l} \rangle A_{l}^{m} =
  F_{l}^{m} \int_{0}^{r_{c}} W_{l}^{m} \left( r \right) 
  \phi^{2} \left( r \right) dr \;.
\end{equation}
$F_{l}^{m}$ is a prefactor of the real spherical harmonics $Z_{l}^{m}$.
Its explicit expressions can be found in, e.g., 
Ref.~\onlinecite{RiOpEsJo1992}. 
$W_{l}^{m}$ is the effective potential at the Nd site expanded by $Z_{l}^{m}$. 
$\phi$ is the radial function of the Nd-$4f$ orbital, 
which is obtained in GGA with the self interaction correction. 
We use a cutoff radius $r_{c}$ as treated in Ref.~\onlinecite{HaTeKiIsMi2015c}.

%%% Results and Discussion %%%
\section{Results and Discussion} \label{sec:results-discussion}
\begin{figure}[ht]
  \includegraphics[width=\hsize]{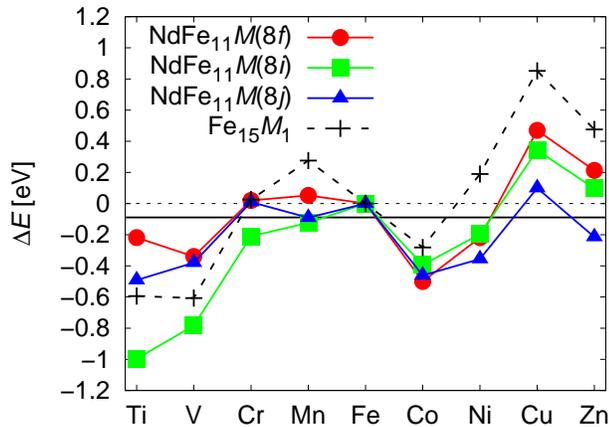}
  \caption{(Color online)
    The formation energy of NdFe$_{11}M$ 
    defined by Eqs.~(\ref{eq:formationenergy_0}) and (\ref{eq:referenceenergy_0}).
    The values corresponding to the substitutional sites, 8$f$, 8$i$, and 8$j$
    are shown as red circles, green squares, and blue triangles, respectively.
    The formation energy of Fe$_{15}M_{1}$ calculated with a fixed bcc-Fe lattice
    is also shown as black crosses.
    The lines connecting the data points are guides for the eyes.}
  \label{fig:formationenergy_3d}
\end{figure}
\begin{figure}[ht]
  \includegraphics[width=\hsize]{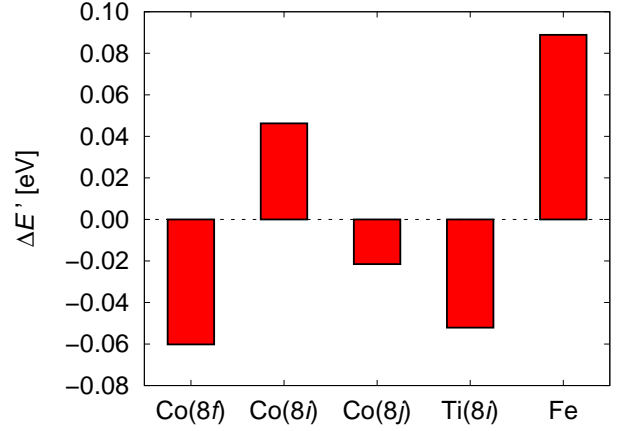}
  \caption{(Color online)
    The formation energy of NdFe$_{11}$Co estimated from
    the total energy for Nd$_{2}$Fe$_{17}$, bcc-Fe,
    and CoFe alloy which has CsCl-type crystal structure
    defined by Eqs.~(\ref{eq:formationenergy_1}) and (\ref{eq:referenceenergy_1}).
    The results for NdFe$_{11}$Ti with Ti substituted at 8$i$ site 
    and for NdFe$_{12}$ are shown for comparison.}
  \label{fig:formationenergy_co}
\end{figure}
\begin{figure}[ht]
  \includegraphics[width=\hsize]{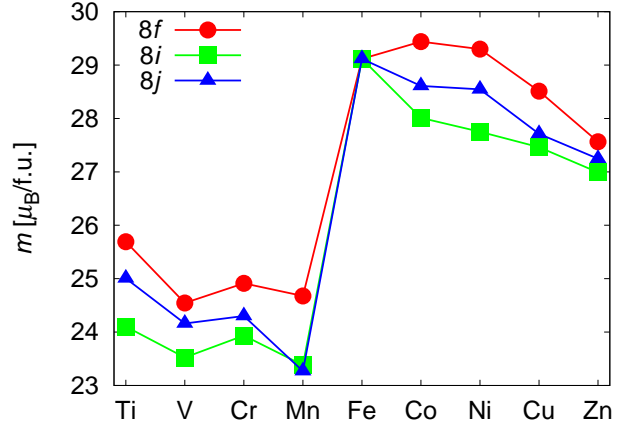}
  \caption{(Color online)
    The total magnetic moment $m$ of NdFe$_{11}M$ described in $\mu_{B}$/f.u.
    The values corresponding to the substitutional sites, 8$f$, 8$i$, and 8$j$
    are shown as red circles, green squares, and blue triangles, respectively.
    The lines connecting the data points are guides for the eyes.}
  \label{fig:spin_3d}
\end{figure}
\begin{figure}[ht]
  \includegraphics[width=\hsize]{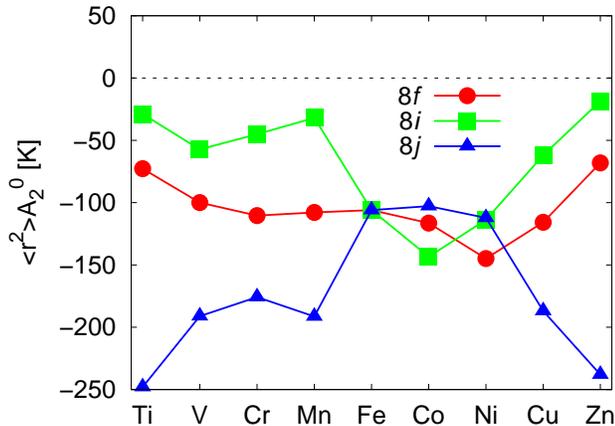}
  \caption{(Color online)
    $\langle r^{2} \rangle A_{2}^{0}$ of NdFe$_{11}M$ in units of K.
    The values corresponding to the substitutional sites, 8$f$, 8$i$, and 8$j$
    are shown as red circles, green squares, and blue triangles, respectively.
    The lines connecting the data points are guides for the eyes.}
  \label{fig:a20_3d}
\end{figure}
\begin{figure}[ht]
  \includegraphics[width=\hsize]{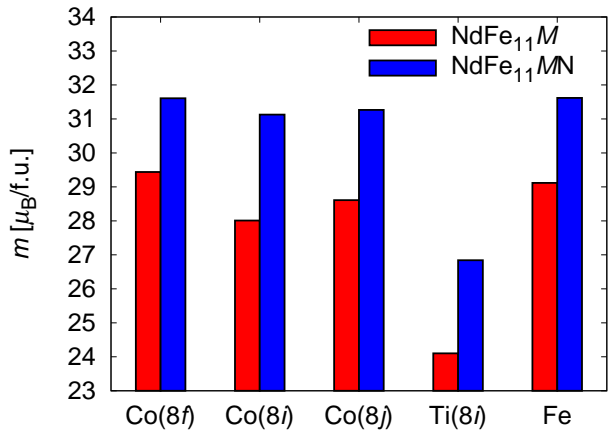}
  \caption{(Color online)
    The total magnetic moment $m$ of NdFe$_{11}$Co (red) and NdFe$_{11}$CoN (blue).
    As a reference, 
    the results for the case of Ti substituted at 8$i$ site and
    for the case without substitution are also shown.}
  \label{fig:spin_co}
\end{figure}
\begin{figure}[ht]
  \includegraphics[width=\hsize]{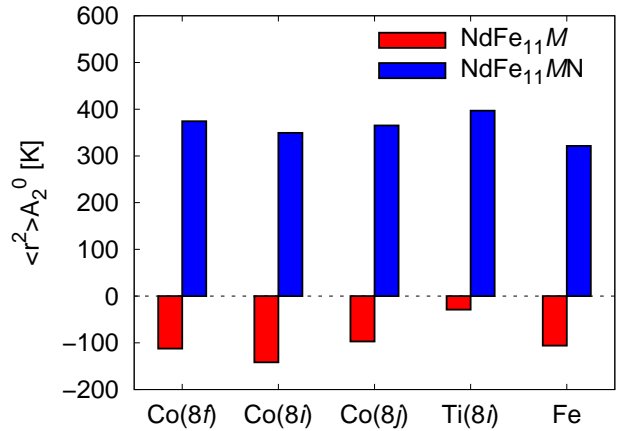}
  \caption{(Color online)
    The $\langle r^{2} \rangle A_{2}^{0}$ of NdFe$_{11}$Co (red) and
    NdFe$_{11}$CoN (blue).
    As a reference, 
    the results for the case of Ti substituted at 8$i$ site and
    for the case without substitution are also shown.}
  \label{fig:a20_co}
\end{figure}

First, we discuss stability by substitution in NdFe$_{11}M$.
We start from defining a formation energy to measure 
how much the substitution stabilizes NdFe$_{11}M$ compared from NdFe$_{12}$,
{\setlength\arraycolsep{2pt}
  \begin{eqnarray}
    \Delta E[M]
    & \equiv & E[\mbox{NdFe}_{11}M] - E_{\text{ref}}[M],
    \label{eq:formationenergy_0}
    \\
    E_{\text{ref}}[M]
    & \equiv & E[\mbox{NdFe}_{12}] - E[\mbox{Fe}] + E[M],
    \label{eq:referenceenergy_0}
  \end{eqnarray}
}
where $E[\mbox{NdFe}_{11}M]$, $E[M]$, and $E[\mbox{Fe}]$
are the total energy of NdFe$_{11}M$ per formula unit, 
the simple substance of $M$ per atom, and bcc-Fe per atom.
%% with their optimized geometries, respectively.
The simple substances of $M$ are chosen as,
hcp (Ti), bcc (V), bcc (Cr), fcc (Mn), hcp (Co), fcc (Ni), fcc (Cu), hcp (Zn).
The lattice constants and inner coordinates are optimized numerically.

Figure~\ref{fig:formationenergy_3d} shows 
the substitutional element dependence of Eq.~(\ref{eq:formationenergy_0}). 
The light transition metal elements, Ti, V, Cr, and Mn has 
negative formation energy. 
Among them, Ti exhibits the largest formation energy in magnitude. 
The preferable site is the 8$i$. 
The 8$j$ site substitution is next preferable,
and 8$f$ is the most unstable among the three sites.
This is consistent with the experimental observation.\cite{Bu1991a}
The order of the preferable sites can be explained by
the canonical bond length as discussed
in Appendix~\ref{app:canonicalbondlength}.
As the atomic number increases,
the formation energy for the most preferable 8$i$ site substitution 
becomes less negative. 
In experiments, the lowest substitution content for phase stability 
increases in the order of Ti-V-Cr-Mn.
This implies that 
the stabilization becomes weak as the atomic number increases, 
which is consistent with the order of the calculated formation energy.

As for the right hand side of Fe in the periodic table, 
Co substitution yields the largest negative formation energy. 
The most stable site for Co is 8$f$, and the next favorable site is 8$j$. 
The two sites have similar formation energy.
Both NdFe$_{11}$Ni and NdFe$_{11}$Zn also exhibit negative formation energy,
but they are less negative than that of Co substitution.

We also plot the calculated formation energy of Fe$_{15}M_{1}$ in which
$M$ is embedded on the 2$\times$2$\times$2 bcc-Fe conventional cell,
\begin{equation} \label{eq:formationenergy_fe-m}
  \Delta E[\mbox{Fe}_{15}M_{1}] \equiv 
  E[\mbox{Fe}_{15}M_{1}] - (15E[\mbox{Fe}] + E[M]).
\end{equation}
The lattice for Fe$_{15}M_{1}$ is fixed to the optimized bcc-Fe structure.
The curve is similar to the formation energy of NdFe$_{11}M$.
NdFe$_{11}M$ contains a lot of Fe and 
$M$ dependence of the formation energy 
can be roughly determined from the coupling of $M$-Fe.
In experiment, the soluble range of NdFe$_{12-x}M_{x}$ depends on 
the substitutional elements.
The experimentally observed range of $x$ for each $M$ may reflect 
the soluble range for the $M$-Fe alloys.

The horizontal black solid line in Fig.~\ref{fig:formationenergy_3d} indicates
a formation energy of Nd$_{2}$Fe$_{17}$ measured from NdFe$_{12}$ 
as a reference system.
\begin{equation} \label{eq:formationenergy_2-17}
  \dfrac{1}{2}\Delta E[\mbox{Nd}_{2}\mbox{Fe}_{17}] \equiv 
  \dfrac{1}{2}E[\mbox{Nd}_{2}\mbox{Fe}_{17}] + \dfrac{7}{2}E[\mbox{Fe}] - E[\mbox{NdFe}_{12}]
\end{equation}
The calculated value of Eq.~(\ref{eq:formationenergy_2-17}) is negative.
Namely, NdFe$_{12}$ is unstable compared to Nd$_{2}$Fe$_{17}$ and bcc-Fe. 
This is consistent with experimental observation. 
By shifting the energy origin from the black broken line 
to the black solid line in Fig.~\ref{fig:formationenergy_3d},
we can read the stability of NdFe$_{11}M$ compared from Nd$_{2}$Fe$_{17}$.

For $M$=Ti, V, and Co, the formation energy defined by 
Eqs.~(\ref{eq:formationenergy_0}) and (\ref{eq:referenceenergy_0})
are large negative values. 
Among these elements, Ti and V substantially reduce the magnetic moment, 
while Co does not (later we will describe in detail).
In terms of both the formation energy and the magnetic moment, 
Co is a good candidate as the substitutional element.

The formation energy defined in Eqs.~(\ref{eq:formationenergy_0}) and (\ref{eq:referenceenergy_0})
does not take into account alloying effects.
Here, we discuss the alloying effect of Co-Fe for the case of $M$=Co.
We redefine the formation energy of NdFe$_{11}M$ comparing from 
Nd$_{2}$Fe$_{17}$ and a $M$-Fe alloy.
{\setlength\arraycolsep{2pt}
  \begin{eqnarray}
    \Delta E'[M]
    & \equiv & E[\mbox{NdFe}_{11}M] - E_{\text{ref}}'[M]
    \label{eq:formationenergy_1}
    \\
    E_{\text{ref}}'[M]
    & \equiv & \dfrac{1}{2}E[\mbox{Nd}_{2}\mbox{Fe}_{17}] 
    + E[M\mbox{Fe}_{x}] + (\dfrac{5}{2}-x)E[\mbox{Fe}] 
    \label{eq:referenceenergy_1}
  \end{eqnarray}
}
$E[M\mbox{Fe}_{x}]$ is the total energy of $M$Fe$_{x}$. 
Figure~\ref{fig:formationenergy_co} shows the calculated formation energy 
defined by Eqs.~(\ref{eq:formationenergy_1}) and (\ref{eq:referenceenergy_1}).
We consider CoFe with the CsCl structure as the reference alloy. 
For comparison,
the results for NdFe$_{11}$Ti (substituted at 8$i$ site) and
NdFe$_{12}$ are also presented. 
For $M$=Ti, TiFe$_{2}$ with MgZn$_{2}$ structure is calculated as the reference.
The calculated formation energy reproduces
the experimental observation that
synthesis of NdFe$_{12}$ is difficult, while that of NdFe$_{11}$Ti is easy. 
NdFe$_{11}$Co for 8$f$ site substitution has negative formation energy
which is as low as that of Ti substitution.
Co can stabilize the ThMn$_{12}$ structure compared from 
the segregation with NdFe$_{12}$, Nd$_{2}$Fe$_{17}$, bcc-Fe, hcp-Co, and CoFe.

The substitution affects not only the stability, 
but also the magnetism of NdFe$_{11}M$. 
Figure~\ref{fig:spin_3d} shows $M$ dependence of the magnetic moments. 
NdFe$_{11}M$ with $M$=Ti, V, Cr, and Mn have 
much smaller magnetic moments compared to NdFe$_{12}$. 
There is a jump in the total magnetic moment 
between the light transition metals Ti--Mn and 
the heavy transition metals Fe--Zn. 
The jump is the result that the magnetic coupling between $M$ and 
the host Fe is antiferromagnetic for these light transition metals and 
becomes ferromagnetic for these heavy transition metals
(for Cu and Zn, it is marginal, i.e., their local moments are almost zero).

The change of the magnetic moments from $M$=Ti to Mn 
is qualitatively explained in terms of Friedel's concept of 
virtual bound state~\cite{Fr1958} by noticing that 
the majority spin $d$ band of NdFe$_{12}$ is nearly filled like in fcc-Ni. 
By substituting $M$(=Ti--Mn), 
the impurity level for the majority spin states appears above 
the top of the host Fe $d$ band and the Fermi level.
The change of the number of the majority spin electrons $\Delta n^{\uparrow}$ is $-5$.
For the minority spin states, 
the host Fe $d$ band and the impurity level are fairly close 
and these states can hybridize.
The electrons introduced by the substitutional element occupy these hybridized states.
The change of the number of the minority spin electrons $\Delta n^{\downarrow}$ is 
$Z(M) - Z_{\text{Fe}} - \Delta n^{\uparrow}$, 
where $Z_{\text{Fe}}$ is the number of the valence electrons of Fe ($=8$),
and $Z(M)$ is those of $M$, i.e., 4 (Ti), 5 (V), 6 (Cr), 7 (Mn).
Then the change in the magnetic moment is given by 
$\Delta n^{\uparrow}-\Delta n^{\downarrow}=-(Z(M)+2.0)$ $\mu_{\mathrm{B}}$.
The deviation from idealistic Friedel's concept is due not only to the fact that 
the majority spin $d$ band of NdFe$_{12}$ is not completely filled 
but also to the presence of continuous $sp$ band overlapping the $d$ band. 
For Mn, the split-off Mn $d$ states are rather close to the Fermi level and 
the significant portion of their tails are occupied to reduce 
the magnetic moment reduction.

Co substitution at the preferable site (the tight 8$f$ site) 
works positively to the magnetic moment. 
For more spacious 8$i$ and 8$j$ sites, on the other hand, 
the magnetic moment is slightly smaller than that of NdFe$_{12}$.
This is in contrast with naive expectation from the Slater-Pauling curve that 
the magnetic moment is increased by slight amount of Co doping to bcc-Fe. 
In Co doped bcc-Fe, 
the electronic structure at the Fe sites next to Co is modified 
due to the upward shift of the minority spin $d$ level of Fe 
caused by the hybridization with the neighboring Co. 
The minority spin occupation decreases due to this level shift 
and the majority spin occupation increases due to the reduction 
in the Coulomb repulsion from the minority spin electrons. 
Therefore, the magnetic moment of Fe next to Co increases slightly. 
The small increase of the magnetic moment at 8 Fe sites over-cancels 
the reduction of the magnetic moment at Co site. 
In the present NdFe$_{12}$ case, 
we see similar behavior at the tight 8$f$ site. 
However, for more spacious sites (8$i$ and 8$j$), 
as the hybridization between Fe $d$ states and Co $d$ states will become weaker, 
the magnetic moment enhancement at the neighboring Fe sites will be reduced. 
Another possible reason for the difficulty of magnetic moment enhancement 
due to Co substitution in NdFe$_{12}$ case may be its smaller bandwidth and 
consequent nearly filled majority spin $d$ band.~\cite{HaTeKiIsMi2015c}
As the atomic number increases further from Co to Zn, 
the magnetic moment decreases monotonically.

Next we discuss the effect of substitution on
the crystal field parameter $\langle r^{2} \rangle A_{2}^{0}$.
The substitutional element dependence of $\langle r^{2} \rangle A_{2}^{0}$
is shown in Fig.~\ref{fig:a20_3d}.
For $M$=Co, 8$f$ site substitution gives only slightly smaller
and 8$j$ site substitution exhibits
almost the same $\langle r^{2} \rangle A_{2}^{0}$
compared to NdFe$_{12}$.
Co substitution at the 8$i$ site decreases $\langle r^{2} \rangle A_{2}^{0}$
by about 50 K.
While Co or Ni substitution brings small change ($<$ 50K) 
irrespective of the substitutional site, 
$\langle r^{2} \rangle A_{2}^{0}$ for other elements 
depends strongly on the substitutional site, 
and the value increases in the order of 8$j$-8$f$-8$i$.
This trend is desirable for the light transition metals, 
since the preferable site is 8$i$.
On the other hand, the preferable site for Cu and Zn is 8$j$, 
hence substitution with Cu or Zn is expected to be unfavorable 
in terms of the magnetocrystalline anisotropy. 

The difference in $\langle r^{2} \rangle A_{2}^{0}$ between the substitutional sites %ex.) for Ti, 
can be explained by the electron density around the Nd site.
%In a previous study\cite{MiTeHaKiIs2014}, 
%the electron charge difference due to the substitution is analyzed 
%to explain the enhancement in $\langle r^{2} \rangle A_{2}^{0}$ with Ti at 8$i$ site.
%We analyze the charge difference as in the previous study.
%We found that 
A Ti atom tends to repulse the electron density from the region between Nd and Ti. 
%positive charge arises in this region. %along the direction around the Nd site.
Hence, if a Ti occupies the 8$i$ site, 
positive charge is accumulated along $a$ (or $b$) axis in the vicinity of the Nd site.
Meanwhile, the positive charge is induced in the $c$ direction by Ti(8$j$), 
and the situation is in between in the case of Ti(8$f$). 
For Co, the things are the other way around, 
and the charge difference is rather small.
These positive charge density attracts the Nd-4$f$ electron charge.
The positive charge appearing, for example, along $a$ axis
enhances uniaxial anisotropy for the Nd-4$f$ moment.
This explains the site dependence of $\langle r^{2} \rangle A_{2}^{0}$.

Finally, we study the effects of nitrogenation for the good candidate, Co. 
Figure~\ref{fig:spin_co} shows the total magnetic moments of \NdFeCo\ (red), 
and their nitrogenated compounds (blue).
The results for Ti substituted at 8$i$ site and 
for the case without substitution are also shown as references. 
The total magnetic moments of \NdFeCo\ and their nitrogenated compounds 
have almost the same values as those of \NdFe\ and 
its nitrogenated compound, 
and have much larger values than those of \NdFeTi(8$i$), 
which are previously studied in theories and experiments.
\cite{MiTeHaKiIs2014,HaTeKiIsMi2015c,YaZhKoPaGe1991,YaZhGePaKoLiYaZhDiYe1991,HiTaHiHo2015}

Next the values of $\langle r^{2} \rangle A_{2}^{0}$ are shown in Fig.~\ref{fig:a20_co} 
for the compounds corresponding to those in Fig.~\ref{fig:spin_co}. 
\NdFeM\ have negative values, 
but their nitrogenated compounds show large positive values as expected. 
The values of \NdFeCoN\ is slightly larger than that of \NdFeN\ and 
are comparable to the values of \NdFeTiN.
These results suggest that Co works positively in \NdFeMN,
therefore possibly yields another good candidate of 
high performance bulk magnet materials.

%%% Conclusion %%%
\section{Conclusion} \label{sec:conclusion}
We have performed first-principles calculations of \NdFeM\ 
where $M$=Ti, V, Cr, Mn, Fe, Co, Ni, Cu, Zn, and 
a few of their nitrogenated compounds. 
The calculated formation energy of \NdFeCo\ is negative 
compared to Nd$_{2}$Fe$_{17}$, bcc-Fe and CoFe in the CsCl structure, 
%% suggesting that \NdFeCo\ could be stable in the bulk.
suggesting that Co could stabilize the ThMn$_{12}$ structure in the bulk.
Furthermore, its nitrogenated compound, 
\NdFeCoN\ enhances the magnetic moment, 
and has almost the same magnetic moment as that of \NdFeN\ 
which is unstable in the bulk.
The magnetocrystalline anisotropy of \NdFeCoN\ estimated from 
the crystal field parameter $\langle r^{2} \rangle A_{2}^{0}$ is almost the same 
compared to \NdFeN\ and \NdFeTiN. 
These data conclude that Co could be a good substitutional element
in \NdFeMN.

We here note that the phases with non-stoichiometric substitutions, e.g.  
NdFe$_{12-x}M_{x}$, Nd$_{2}$Fe$_{17-x}M_{x}$, 
are not discussed in our analysis.
To improve performance, 
one possible strategy is to reduce Ti content $y$ in NdFe$_{12-x-y}M_{x}$Ti$_{y}$. 
The result for the formation energy indicates that
$M$=Co substitution enables it.
This point could be important and remains as a future work.

%%% Acknowledgement %%%
\begin{acknowledgments}
The authors would like to thank 
Dr. S. Hirosawa for fruitful discussions.
This work was supported by 
the Elements Strategy Initiative Project under the auspice of MEXT, 
by “Materials research by Information Integration”
Initiative (MI$^2$I) project of the Support Program for Starting Up Innovation Hub from
Japan Science and Technology Agency (JST),
and also by 
MEXT as a social and scientific priority issue 
(Creation of new functional Devices and high-performance Materials to 
Support next-generation Industries; CDMSI) to be tackled by using post-K computer.
The computation has been partly carried out using the facilities of 
the Supercomputer Center, the Institute for Solid State Physics, 
the University of Tokyo, and the supercomputer of ACCMS, Kyoto University, and 
also by the K computer provided by the RIKEN Advanced Institute for 
Computational Science (Project ID:hp140150, hp150014 and hp160227). 
\end{acknowledgments}

\appendix
\section{Space of Fe sites in NdFe$_{12}$} \label{app:canonicalbondlength}
\begin{figure}[ht]
  \includegraphics[width=\hsize]{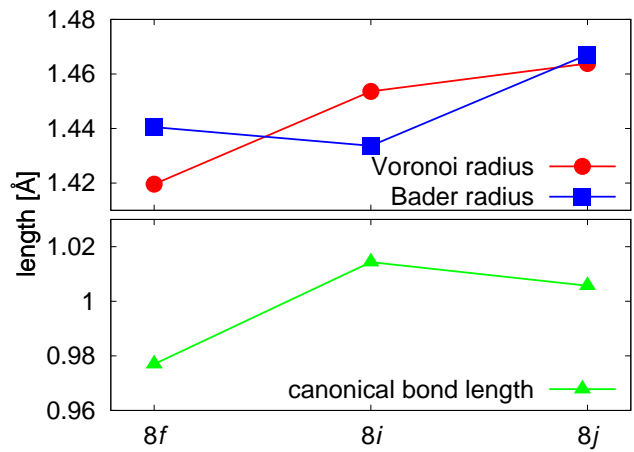}
  \caption{(Color online)
    The upper figure shows
    the lengths estimated from Voronoi cell volume and from Bader cell volume, respectively.
    The lower figure shows the canonical bond length 
    defined in Eq.~(\ref{eq:length_canonicalbandtheory}). 
    The lines connecting the data points are guides for the eyes.}
  \label{fig:bondinglength}
\end{figure}

For a complex crystal structure, 
there is an ambiguity for choosing "neighboring" atoms, 
thus, for defining the average atomic distance.
In this appendix, 
we propose a length scale to describe effective sparseness 
reflecting strength of the hybridization. 
This length scale can be useful to discuss 
the preferential substitutional sites in NdFe$_{11}M$. 
In addition, the local spin moment can be also discussed using this length.

In NdFe$_{12}$, there are three Fe sites, 8$f$, 8$i$, and 8$j$.
We expect that a preferable substitutional site is explained 
by the space of each site.
To compare the space of atomic site, we introduce 
a length $l_{\mu}$ based on the canonical band theory.\cite{An1975}
\begin{equation} \label{eq:length_canonicalbandtheory}
  l_{\mu} \equiv \dfrac{1}{2}\left(
  \sum_{\nu} \left|\Vec{r}_{\nu}-\Vec{r}_{\mu}\right|^{-10}
  \right)^{-\frac{1}{10}}
\end{equation}
where $\mu$ specifies the atomic sites, 8$f$, 8$i$, and 8$j$.
The summation is taken over the neighboring atoms.
The canonical bond length $l_{\mu}$ defined above is a measure of the $d$-$d$ bond.
From the canonical band theory, 
the $d$-$d$ hopping integral is proportional to inverse 5th power of the distance
between the $d$ orbitals. 
The bond strength is measured in terms of the width of the $d$ band 
which is given by a square root of the sum of hopping integral squared 
(inverse 10th power of the distance) over the neighboring sites of a given site.

Figure~\ref{fig:bondinglength} shows $l_{\mu}$ 
for Fe-Fe bonds in NdFe$_{12}$. 
For comparison, we also plot the Voronoi radius and Bader radius.
The Voronoi and Bader radii are determined 
so that the volume of the spheres defined by these radii are equal to
the volumes of the Voronoi cell and Bader cell, respectively.
The Voronoi cell of a $\mu$th atom is defined as 
a region surrounded by the perpendicular bisection planer boundary 
between the $\mu$th atom and its neighbors 
(see Ref.~\onlinecite{Be1988} for the details of the calculation).
The Bader cell is given by the space decomposition according to 
the Bader population analysis\cite{Ba1990,YuTr2011} 
for sum of the partial electronic core density, the pseudo charge density, 
and the compensation charge density 
(their notations are following Ref.~\onlinecite{KrJo1999}).
Light transition metals such as Ti, 
which has larger atomic radius than Fe,
occupies the 8$i$ site. 
Hence, we naively expect that the 8$i$ site is the most spacious site 
among the 8$f$, 8$i$, and 8$j$ sites. 
The canonical bond length is consistent with this expectation. 
On the other hand, the Voronoi radius does not take 
the maximum value for the 8$i$ site, 
and the Bader radius takes the smallest value at 8$i$. 

The relative magnitude of the local spin moment can be also understood
by using the canonical bond length.
The canonical bond length reflects the strength of the hybridization 
with surrounding atoms, 
i.e., larger (smaller) $l_{\mu}$ corresponds to weaker (stronger) hybridization.
As the hybridization becomes weaker (stronger), 
the band width becomes narrower (broader).
The band width is roughly in inverse proportion to 
the amplitude of the density of states.
Following the Stoner criterion for ferromagnetism, 
larger density of states at Fermi level is expected to give 
stronger spin polarization.
Eventually, 
larger $l_{\mu}$ value is expected to yield stronger spin polarization.
The local spin moments of Fe at 8$f$, 8$i$, and 8$j$ sites are 
1.78, 2.53, and 2.31 $\mu_{B}$, 
which is consistent with the expectation from $l_{\mu}$.

\bibliography{./Reference}

\end{document}